# Cathode depth sensing in CZT detectors

J. Hong[a*], E. C. Bellm[a], J. E. Grindlay[a], and T. Narita[b]

[a]Harvard Smithsonian Center for Astrophysics, 60 Garden St., Cambridge, MA 02138
[b]College of the Holy Cross, 1 College St., Worcester, MA 01610

## ABSTRACT

Measuring the depth of interaction in thick Cadmium-Zinc-Telluride (CZT) detectors allows improved imaging and spectroscopy for hard X-ray imaging above 100 keV. The Energetic X-ray Imaging Survey Telescope (EXIST) will employ relatively thick (5 - 10 mm) CZT detectors, which are required to perform the broad energy-band sky survey. Interaction depth information is needed to correct events to the detector "focal plane" for correct imaging and can be used to improve the energy resolution of the detector at high energies by allowing event-based corrections for incomplete charge collection. Background rejection is also improved by allowing low energy events from the rear and sides of the detector to be rejected.

We present experimental results of interaction depth sensing in a 5 mm thick pixellated Au-contact IMARAD CZT detector. The depth sensing was done by making simultaneous measurements of cathode and anode signals, where the interaction depth at a given energy is proportional to the ratio of cathode/anode signals. We demonstrate how a simple empirical formula describing the event distributions in the cathode/anode signal space can dramatically improve the energy resolution. We also estimate the energy and depth resolution of the detector as a function of the energy and the interaction depth.

We also show a depth-sensing prototype system currently under development for EXIST in which cathode signals from 8, 16 or 32 crystals can be read-out by a small multi-channel ASIC board that is vertically edge-mounted on the cathode electrode along every second CZT crystal boundary. This allows CZT crystals to be tiled contiguously with minimum impact on throughput of incoming photons. The robust packaging is crucial in EXIST, which will employ very large area imaging CZT detector arrays.

**Keywords:** CZT detector, depth sensing, hard X-ray imaging

## 1. INTRODUCTION

Over the past few years, Cadmium-Zinc-Telluride (CZT) has become the most promising material for hard X-ray detectors. Its high atomic number elements efficiently stop X-rays at reasonable thickness. Despite electron trapping and relatively poor hole mobility of CZT crystals, many techniques such as pixellization[1] and coplanar grid[2] have been developed to provide good energy resolution, substantially superior to that of conventional scintillators employing NaI and CsI crystals and even approaching Ge spectrometers for some applications. In particular, pixellization allows excellent positional sensitivity. Operability at room temperature also makes CZT detectors very attractive.

Therefore, CZT detectors are the first choice for detectors in the proposed Einstein Probe Black Hole Finder mission, *Energetic X-ray Imaging Survey Telescope* (EXIST)[3]. The primary goal of EXIST is to find obscured black holes by directly detecting their hard X-ray emission. Obscured black holes in galactic nuclei are largely inaccessible in other lower energy bands. EXIST will also perform a long overdue very sensitive hard X-ray sky imaging sky survey (10 – 600 keV) and it will serve as a next generation Gamma-ray Burst Observatory.

In order to perform wide-field hard X-ray imaging, EXIST will employ the coded-aperture technique, using CZT detectors with Tungsten masks. The broad energy band coverage (10 – 600 keV) requires thick detectors (~ 5 – 10 mm), and high angular resolution (~5 arcmin) to avoid source confusion requires small detector pixels (1.25 mm pitch for 1.5 m separation of masks and detectors). The high sensitivity (~ 0.05 mCrab below ~100 keV) and long duty cycle (≥20%) requirements call for a large area detector (~ 8 m$^2$) and wide field of view (180$^o$ × 75$^o$) for full-sky imaging each orbit.

---

[*] Send correspondence to J. Hong (jaesub@head-cfa.cfa.harvard.edu)

In coded-aperture imaging, when reconstructing sky images from data taken by detectors with a relatively large ratio of thickness to pixel diameter the lack of depth information for X-ray interactions is effectively translated into inaccurate pixel identification, particularly for X-rays coming at a far off-axis angle. The resulting ambiguity in pixel location on the imaging plane degrades the angular resolution of the telescope. Therefore, depth sensing is necessary for thick, small pixel detectors for wide-field imaging.

Besides restoring angular resolution, depth sensing can improve energy resolution by allowing correction for depth-dependent charge collection efficiency or anomalies. Depth sensing also enhances background rejection by singling out certain types of background events such as shield leakage and internal background. With depth sensing, one can operate detectors in Compton telescope mode. Finally depth sensing can improve the polarization sensitivity of detectors if there is also multi-site (multi-pixel) event readout capability.

The depth sensing technique we employ here is so-called cathode depth sensing, which is to acquire depth information of X-ray interactions by simultaneous measurements of cathode and anode signals from the interactions in a pixellated detector. In our detector, the anode contact is pixellated and the cathode side faces X-ray sources due to the complex electronics readout systems closely attached to the anode side. For a given X-ray interaction in the detector, the signal gathered by an anode pixel is proportional to the charge induced by the interaction, while the cathode side collects a signal that is proportional to the induced charge times the depth of the interaction. Therefore, the ratio of two signals is roughly proportional to the depth of the interaction. Cathode depth sensing is now a standard way of extracting depth information of interactions in a pixellated CZT detector.

Primarily motivated by the EXIST mission, we are developing technology for a depth sensing technique that can be easily applied to a large area detector with minimal impact on throughput of incident X-rays. The very large area detector needed would consist of smaller, identical modules. Thus, our interest is to build a depth sensitive detector module allowing tight and efficient packaging into a large area. In this paper, we demonstrate such a technique using the third generation of our CZT detector system, CZT3, with multi-crystal cathode depth sensing capability. Furthermore we present an absolute depth calibration scheme than can be easily applied to a large area detector. Finally we present some of depth sensing data and derived results such as improved spectral resolution we have obtained with a prototype version of our system in order to illustrate some of the benefits from depth sensing.

## 2. DETECTOR SETUP

### 2.1. CZT crystals and anode electronics

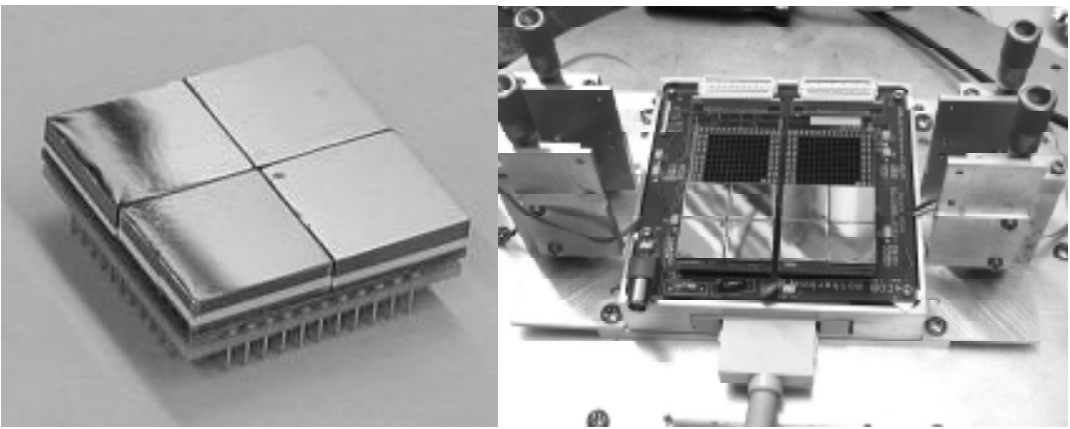

**Figure 1.** 2 × 2 arrays of Au and Pt contact IMARAD CZT crystals in a DCA (left) and the half-filled motherboard of the CZT3 module (right).

We employ 5 mm thick 19.44 × 19.44 mm$^2$ IMARAD CZT crystals, with each crystal having 8 × 8 pixels with 1.9 mm pad size and 2.5 mm pitch. Standard In-contact IMARAD CZT crystals exhibit ohmic behavior (~10$^{10}$ Ω-cm bulk resistivity)[4]. We replace the In-contacts with Au or Pt to have Schottky barrier performance (~10$^{11}$ Ω-cm bulk resistivity)[4]. At a typical bias voltage of -700V, the leakage current for a 2.5 mm pitch pixel of Au- or Pt-contact IMARAD CZT crystals is ~ 1 nA, which is smaller than that of In-contact crystals by about a factor of 10 - 15.

Four of these crystals and the readout electronics are assembled into a unit called DCA (detector crystal array) as shown in Fig. 1. In a DCA, crystals are tiled into a 2x2 array and anode pixels on each crystal are laid out asymmetrically to preserve the pixel pitch across the crystal gaps of the DCA. Underneath the crystals, two 128-channel IDEAS XAIm 3.2 ASICs are mounted through a coupling board, handling signals from 256 anode pixels of the 4 crystals. In the CZT3 system, we can mount four DCAs (16 ×16 cm$^2$, with 0.5 pixel wide gaps between DCAs) on the motherboard (Fig. 1), which is controlled by an XA anode controller box through a PC computer. The XA anode controller processes the anode signals from the 8 XAIm ASICs reading out the 16 crystals.

**2.2. Cathode depth sensing scheme**

Among various approaches to read out the cathode signal, we have developed the following scheme in this study. We mount a cathode electronics board vertically with a micrometer-controlled support structure (Fig. 2). The signal from each crystal is read out through a spring-loaded pogo-pin directly contacting the cathode surface of the crystal. The cathode signals are amplified using an IDEAS VA32_500 ASIC, which is capable of 32 channel parallel processing. The ASIC is mounted on the vertical cathode board through a small carrier board. Every two channels in the ASIC are assigned to each crystal: one for the signal and the other for the reference for the differential output to cancel out electronics noise. In the current configuration, the VA32_500 ASIC handles signals from 8 crystals, but one VA32_500 ASIC can handle as many as 16 crystals in this scheme.

The vertical mounting scheme of cathode electronics boards offers two advantages. First, one can easily achieve a modular system that can be closely tiled into a large area. In the current configuration, the bulky structures such as the micrometer controlled mounting stage, pogo-pin contacts, the ASIC carrier board, etc. are employed mainly because of repetitive mounting/un-mounting of crystals and ASICs for laboratory prototype development and testing. For a flight system, they can be easily replaced by more compact components and conductive epoxy connections. Second, vertically mounted boards present the minimum cross section for incodemt X-rays, which is important in order to keep or extend the lower limit of energy band of detectors below ~10 keV. In addition, EXIST will likely require some form of low energy collimators due to relatively high population of bright X-ray sources in the sky at energies below ~10 keV. Therefore, carefully designed vertical cathode electronics boards can serve such a role as well.

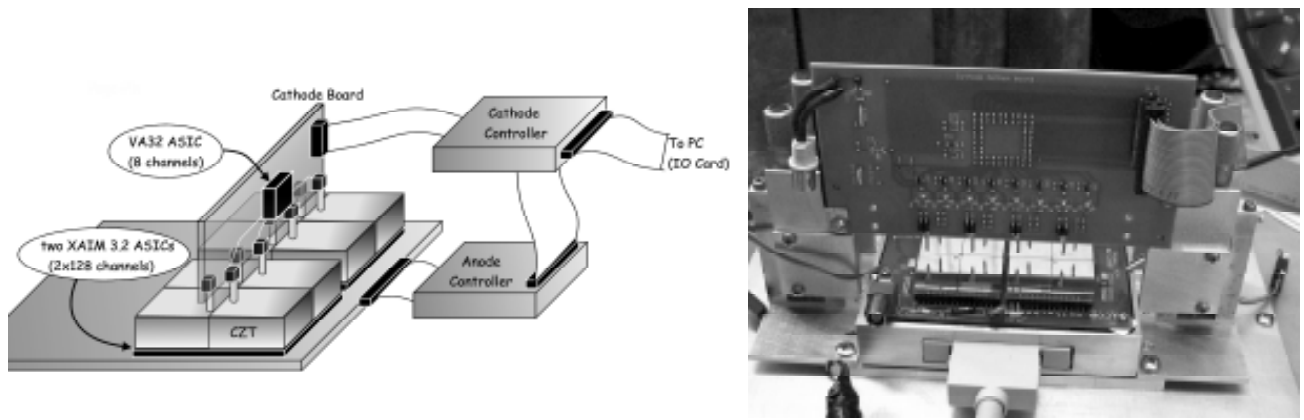

**Figure 2.** Electronics scheme of the CZT3 system capable of multi-crystal cathode depth sensing (left) and the actual picture of a vertical board mounted on the CZT3 system (right). A VA32_500 ASIC mounted on the vertical board (back side of the picture) is used to read out cathode signals from 8 crystals through spring loaded pogo-pins. Each two channels in the ASIC are assigned to each crystal: one for the signal and the other for the reference.

We custom-designed a cathode controller box to process amplified cathode signals and to control the ASIC. Since VA32_500 ASICs continuously read out signals without self-triggering, we use a fast trigger signal from the XA anode controller box (~500 ns before the peak) to sample and hold cathode signals of all 8 crystals. Meanwhile we locate the actual crystal housing the event, using a programmable EPROM (CY7C281) that contains a conversion map from anode pixel addresses to crystal ID. The held signal from the properly selected crystal is then digitized by a 12 bit ADC (AD9220) and all 12 bit data are fed into the computer through an IO card, following the digitized anode signal (12 bit) and the anode pixel address (12 bit). The complete event processing requires ~3-4 μsec.

### 3. ABSOLUTE DEPTH CALIBRATION SCHEME

Although the ratio of cathode to anode signal for a given X-ray energy is roughly proportional to the depth of the interaction, in reality the relation is not quite so simple for various reasons. We need to calibrate cathode detector signals by the known depth of interactions for a given X-ray energy. To do so, we require predetermined depth information of the interactions. One approach to acquire the depth information independently is to limit the range of the depth of interactions in detectors.

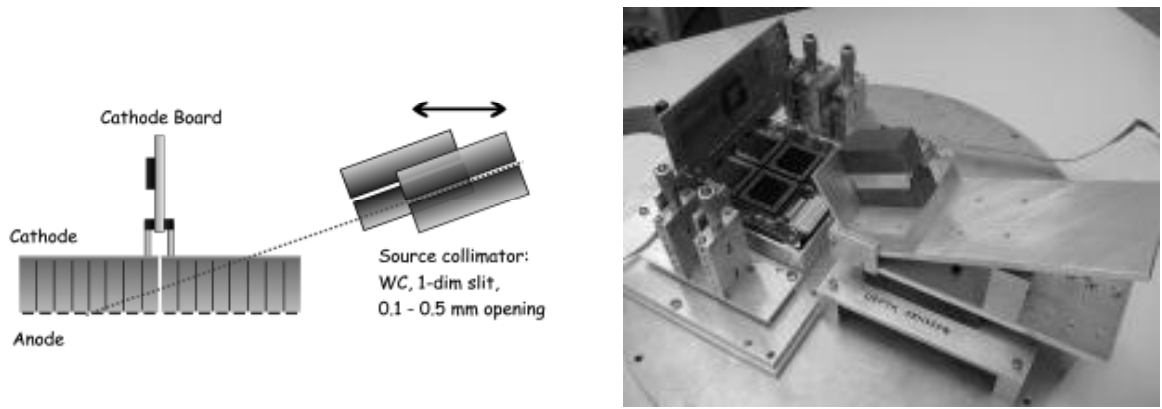

**Figure 3.** The absolute depth calibration scheme (left) and the picture of the source collimator stage (right): two $2 \times 2 \times 5$ cm$^3$ tungsten carbide bars are used to collimate radiation from a source (not shown) into a very narrow (~0.1mm) off-axis fan beam. The collimator mounting stage can set the beam elevation from 10 to 25 degrees or 0 degree (horizontal beam) and the whole stage is mounted on a translation micrometer that can control the precise location of the irradiation point on the crystals.

We set up a far off-axis, narrowly collimated radiation beam as shown in Fig. 3, so that, interactions in each pixel are constrained to occur in a small range of the depth. Radiation from a radioactive source (e.g. $^{133}$Ba) is collimated by a 1-dimensional slit, which consists of two $5 \times 2.5 \times 2.5$ cm$^3$ tungsten carbide slabs. The opening of the slit is set by the separation of two slabs, and the separation is controlled by the number of thin Al layers in between the bars (attached at the edges). This collimator can have an opening of anywhere from 0.1 to 0.5 mm, which sets a beam size of 0.5 – 2.5 mm on the cathode surface of crystals, depending on the beam slope and the relative distance (10 – 20 cm) between the collimator and the crystals. For a narrow beam (~0.1 mm slit opening), interactions in each pixel are confined to a limited range of the depth (~ 0.5 – 1.0 mm).

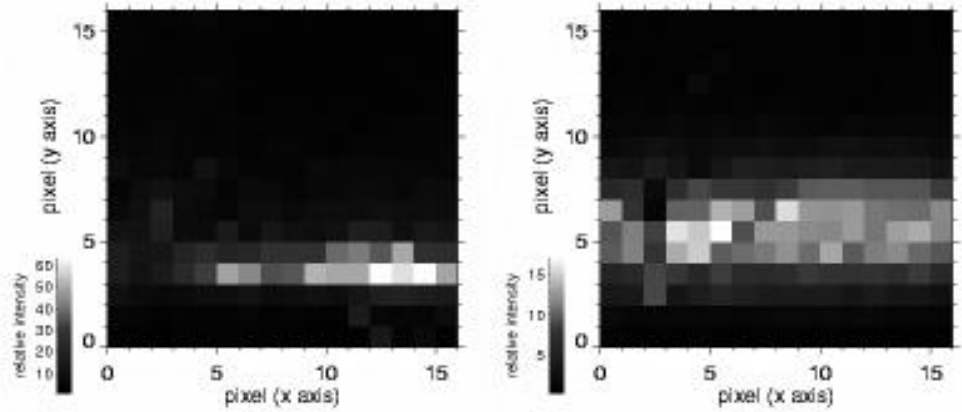

**Figure 4.** Preliminary data for an alignment test of an off-axis radiation beam: count rate map of anode signals in detector space (2 x 2 crystals = 16 pixels × 16 pixels) using a $^{133}$Ba radioactive source with a crude collimation. The left panel shows the low-energy events (30 and 80 keV lines from $^{133}$Ba) indicating the beam size and the entrance point to the crystal. The right panel shows the high-energy events (primarily the 302 and 356 keV lines). Their relatively wide spread along the *y* axis indicates that the high energy X-ray interactions occur in a wide range of depth in the detector.

The whole collimator stage is mounted on a micrometer-controller actuator that can set the precise location of the irradiation point on the crystals. Using the micrometer, one can perform a precise scan along a direction (the arrow in Fig. 3). Such a scan allows us to easily explore the full range of the depth for all the pixels in a large array of detectors. The beam slope can be set at anywhere from 10 to 25 degrees. We can also shine radiation horizontally into the side of the detectors (0 degree) for illumination of the detector pixels at constant depth.

Another advantage of the above scheme is the relatively easy alignment of the setup. Before constructing the above mounting stage, we tested the above scheme using a somewhat crudely configured off-axis radiation beam. The result is shown in Fig. 4. We used a $^{133}$Ba radioactive source that generates a few low energy lines (30 and 80 keV) and high energy lines (276, 302, 356, and 383 keV). Since low energy X-rays mostly interact near the cathode side, their detection indicates the beam size and the irradiation point on the surface of crystals (left panel in Fig. 4), while high energy X-ray data can be used to calibrate detector signals (right panel in Fig. 4). In summary, the movable, far off-axis, narrow radiation beams allow us to calibrate depth dependent signals from a large array of detectors and the positional alignment using low energy X-ray data keeps the procedure simple and reproducible.

## 4. DEPTH SENSING DATA

As described in the previous sections, the CZT3 system capable of multi-crystal depth sensing and the absolute depth-sensing calibration system are almost ready. We are only left with fine-tuning of the cathode electronics for the actual data taking. Here we present some depth sensing data and their analysis using the earlier prototype version of the system: the CZT3 system with a single crystal depth sensing using an AmpTek low noise preamplifier (A-225) and a shaping amplifier (A-206). The electronics scheme is summarized in Fig. 5. The cathode signals from one of the Au-contact IMARAD CZT crystals were measured. The output of the A-206 shaping amplifier is further shaped by a spectroscopy amplifier (ORTEC 452) with ~2 μsec time constant, followed by a sample and hold circuit. The held signal is digitized by a 10 bit ADC (AD573), but only the most significant 6 bits are saved due to limited band width of the spare data stream in the earlier version of the system. The data were taken while a $^{133}$Ba radioactive source was shining simply vertically down onto the detector.

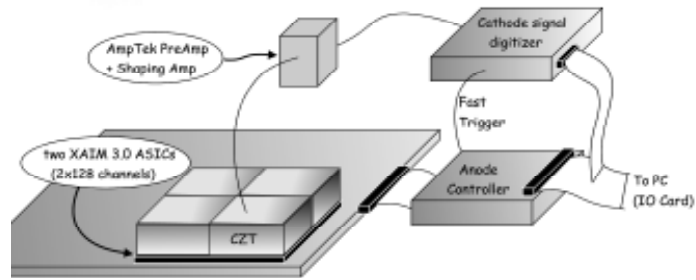

**Figure 5.** The earlier version of the CZT3 system capable of cathode depth sensing on a single crystal: an AmpTek preamplifier (A-225) and a shaping amplifier (A-206) were used to read out cathode signals from a CZT crystal.

### 4.1. Improving energy resolution by depth sensing

For a given line energy source, a typical energy histogram recorded by a pixellated CZT detector has a large low-energy tail, depending on the bias voltage across the detector[1]. The tail is usually due to depth-dependent charge collection efficiency[1]. Therefore depth information can be used to correct signals or even exclude events if such a correction is not possible. As a result, depth information can enhance the energy resolution of CZT detectors.

The left panel in Fig. 6 shows a scatter plot of cathode and anode signals taken in a single pixel. The $^{133}$Ba radioactive source is again used to generate four high-energy lines (276, 302, 356 and 384 keV), which appear as four distinct lines in the plot. The apparent proximity of the 356 and 384 keV lines are due to a nonlinear response of the XAIM ASIC, which is more outstanding at high energies because of its limited dynamic range ($\leq$ 400 keV). For a given energy, the events occurring deep in the detector near the anode side (and which produce a small cathode signal) show a major digression from the trend of the rest. Using a simple empirical formula, one can remove such a digression. We used the following empirical relation,

$$E = \frac{A P_{anode}}{1 - \exp(-B r) - \exp(-C/r)},$$

where $P_{anode}$ is the cathode signal, $r$ is the ratio of cathode to anode signal, and $A$, $B$ and $C$ are constant parameters[1]. We introduce an additional term with $C$, which can correct distortion due to events occurring near the cathode side[5]. The parameters $A$, $B$ and $C$ are determined by fits to the data for several line energies, and are generally different for each pixel since they depend on local values of charge transport in the crystal.

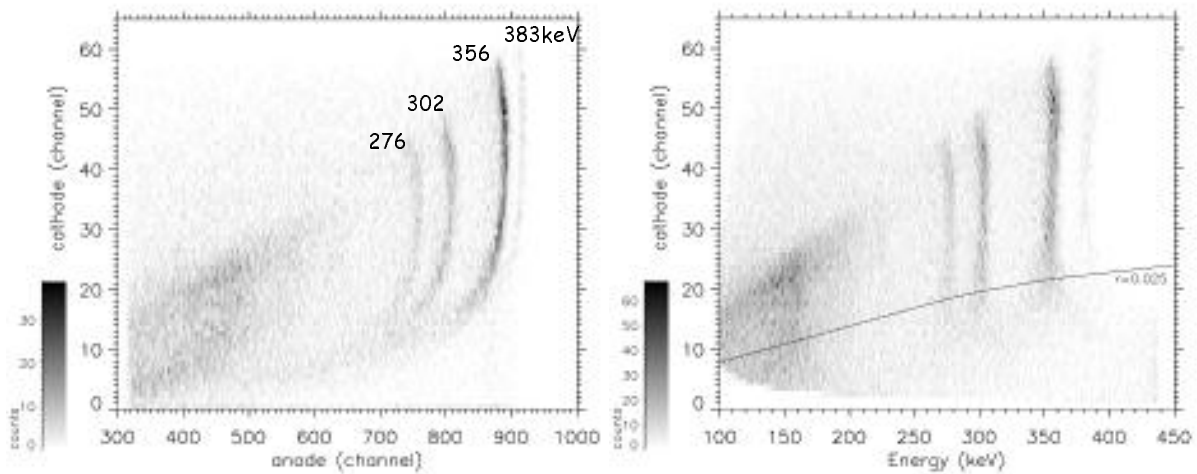

**Figure 6.** Scatter plots of events in raw cathode vs. anode signal (left) and in cathode vs. corrected energy (right)

The cathode signal of the events is plotted against their corrected energy value in the right panel of Fig. 6. The improvement is evident, although there is room for further correction (the "lines" are still not vertical). For events occurring very close to the anode side (events below the $r$=0.025 line in the figure), it is extremely difficult to identify their true energy of the original X-rays. For energy spectra, it is sometimes good enough to simply exclude these events. Fig. 7 summarizes the improvement in energy resolution by this simple correction on the above data. The black histogram shows the uncorrected energy histogram acquired solely from the anode signals (4.3% FWHM at 356 keV), and the orange histogram ($r$>0.025) shows the corrected energy histogram using cathode and anode signals (2.5% FWHM at 356 keV). The blue histogram (lowest curve) shows the result of correcting the remaining (~15%) of the events with r < 0.025; ignoring these (for spectroscopy) would thus reduce the overall detector efficiency by only <15%. This example shows ~ 40% improvement in energy resolution.

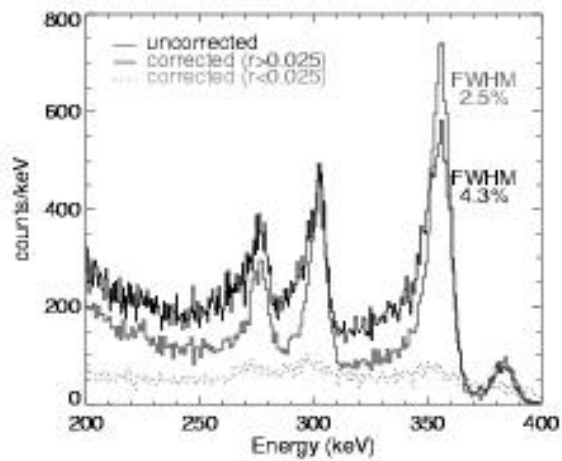

**Figure 7.** Energy histograms before and after the correction by the cathode signal.

### 4.2. Performance estimation of the depth sensing

Based on the cathode depth sensing data, one can estimate the depth and energy sensitivity of the system as a function of the interaction depth and the energy of incident X-rays. Assuming a normal distribution of cathode and anode signals with a proper dispersion based on the pulsar data, first, we randomly generate N cathode and anode signals for a given energy of X-rays interacting at a given depth. Then we use the above formula to convert simulated cathode and anode signals to the estimated values of the interaction depth and the X-ray energy, which we can compare with the given true depth and energy.

Fig. 8 shows an example of such calculations from the simulated events (N=2000 events for each depth and X-ray energy). The left panel shows 95% accurate depth estimation as a function of depth and energy. On the right, the 1σ standard deviation of the energy estimates are shown in the same space. Although some of the assumptions are somewhat naive in producing these results, we expect sub-mm depth resolution and ~1% level energy resolution at high energies (> 200 keV) from depth sensing. These estimations will be readily verified by new measurements using the absolute depth calibration scheme described in section 3 above.

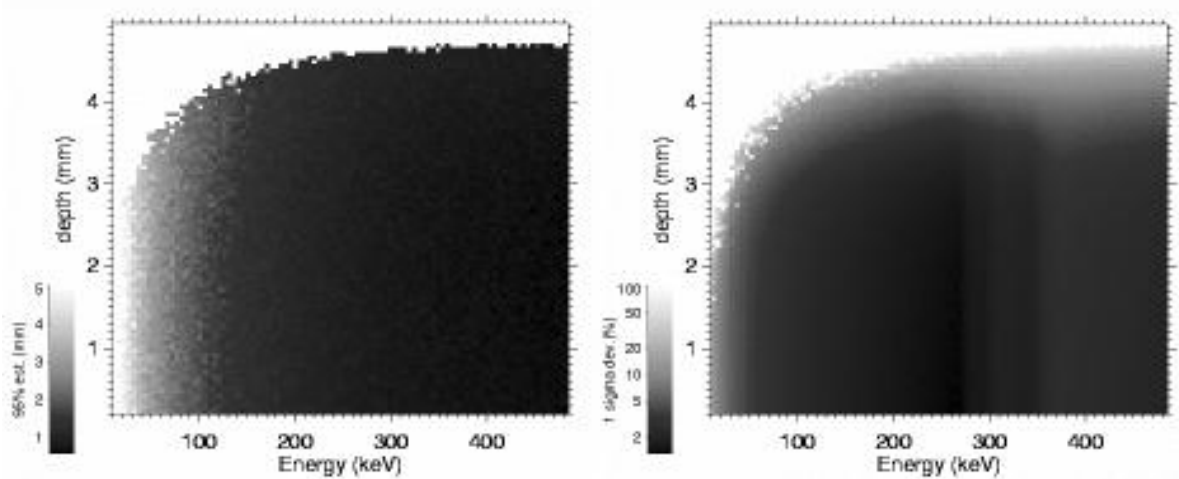

**Figure 8.** Semi-empirical estimations of depth sensitivity (left, 95% est.) and energy resolution (right, 1σ deviations) as a function of the interaction depth and the energy of incident X-rays.

## 5. CONCLUSION

We have shown a simple, multi-crystal, cathode depth-sensing scheme that can be applied to pixellated CZT detector arrays of a large area. We have also introduced a simple technique to efficiently calibrate detector signals to the depth of X-rays interactions over large arrays of CZT detectors. The current depth-sensing scheme can be easily upgraded into a system that allows tight and efficient packaging (Fig. 9). The profile of vertical cathode electronics boards can be easily kept down to provide the required low energy collimation. All the cathode readout electronics can sit on the vertical electronics boards and inside of the surrounding support structures. We plan to develop a few such detector modules (each ~ 256 $cm^2$), and we are going to test the detectors in a near space environment on a balloon flight. We are also studying alternative approaches such as anode depth sensing[6] to get improved depth information in pixellated CZT detectors for the bottom ~20% of the detector nearest the anodes.

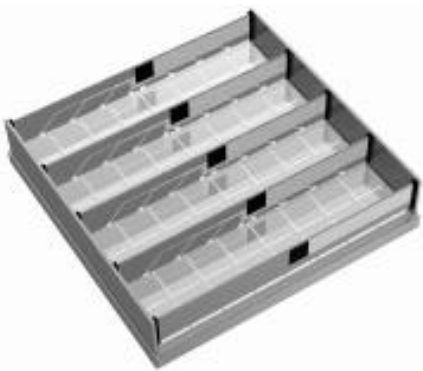

**Figure 9.** The next proto-type detector module concept for a balloon flight: the vertical board will have a small profile and all the necessary electronics willl reside on the vertical board and inside of the supporting structures. An ASIC (black square) will handle cathode signals from 16 crystals.

## ACKNOWLEDGMENTS

This work is supported in part by NASA grant NAG5-5279.